# Ultrafast intrinsic optical-to-electrical conversion dynamics in graphene photodetector

Katsumasa Yoshioka[1,*], Taro Wakamura[1], Masayuki Hashisaka[1], Kenji Watanabe[2], Takashi Taniguchi[3], and Norio Kumada[1]

[1]NTT Basic Research Laboratories, NTT Corporation, 3-1 Morinosato-Wakamiya, Atsugi, 243-0198, Japan

[2]Research Center for Functional Materials, National Institute for Materials Science, 1-1 Namiki, Tsukuba 305-0044, Japan

[3]International Center for Materials Nanoarchitectonics, National Institute for Materials Science, 1-1 Namiki, Tsukuba 305-0044, Japan

*e-mail: katsumasa.yoshioka.ch@hco.ntt.co.jp

**Abstract: Optical-to-electrical (O-E) conversion in graphene is a central phenomenon for realizing anticipated ultrafast and low-power-consumption information technologies. However, revealing its mechanism and intrinsic time scale require uncharted terahertz (THz) electronics and device architectures. Here, we succeeded in resolving O-E conversion processes in high-quality graphene by on-chip electrical readout of ultrafast photothermoelectric current. By suppressing the RC time constant using a resistive zinc oxide top gate, we constructed a gate-tunable graphene photodetector with a bandwidth of up to 220 GHz. By measuring nonlocal photocurrent dynamics, we found that the photocurrent extraction from the electrode is instantaneous without a measurable carrier transit time across several-micrometer-long graphene, following the Shockley-Ramo theorem. The time for photocurrent generation is exceptionally tunable from immediate to > 4 ps, and its origin is identified as Fermi-level-dependent intraband carrier-carrier scattering. Our results bridge the gap between ultrafast optical science and device engineering, accelerating ultrafast graphene optoelectronic applications.**

## Main text:

As data traffic is expected to continue its exponential growth, there is an urgent demand for ultrahigh-bandwidth and low-power-consumption optical receivers, which convert an optical signal into an electrical one[1,2]. Photothermoelectric (PTE) graphene photodetectors (PDs)[2–9] are promising platforms for optical-to-electrical (O-E) conversion thanks to their zero dark current operation[5–7], broadband absorption[8,9], and high conversion efficiency by hot carrier multiplication (HCM)[10,11]. Femtosecond optical pump-probe measurements[5,10,12–24] suggest that photo-excited nonequilibrium carriers instantaneously thermalize into a Fermi-Dirac distribution with elevated carrier temperature (< 100 fs)[15,18,22,24], and then the temperature decreases with the picosecond time scale through various phonon interactions[9,14,17,20,23]. Because of this ultrafast energy relaxation, the 3-dB bandwidth of graphene PDs is expected to exceed 200 GHz[2,12,25]. However, despite the considerable efforts to construct ultrafast graphene PDs based on the PTE effect in various configurations[2–9], the measured bandwidth has been limited to around 70 GHz[7] due to the bandwidth limitation of the readout electronics such as oscilloscopes or spectral analyzers. More critically, the large RC time constant[5,9] coming from gate capacitance[26], which is necessary to tune the Fermi level, sets the cut-off frequency below 100 GHz. As a result, the O-E conversion working with its intrinsic time scale has not been achieved, and thus the carrier-extraction mechanism remains unexplored, though both are crucial for designing ultrafast graphene optoelectronic devices[2].

In this work, ultrafast PTE currents extracted from graphene PDs were measured on-chip using a laser-triggered photoconductive (PC) switch [26–33]. The RC time constant was minimized by using highly resistive zinc oxide (ZnO) as a top gate[34,35]. As a result, we succeeded in demonstrating graphene PDs with a bandwidth of up to 220 GHz. This indicates that we have overcome the bandwidth limitations for tracking the O-E conversion with its intrinsic time scale.

For a comprehensive understanding of the O-E conversion mechanism in graphene, we thoroughly investigate nonlocal photocurrent dynamics while tuning the Fermi energy in several hBN-encapsulated graphene samples with different mobilities and channel lengths. After photoexcitation, O-E conversion proceeds in four stages: (1) nonequilibrium photo-excited carriers thermalize via intraband HCM[10,11] with the duration of immediate to ~4 ps, depending on the Fermi level; (2) photovoltage is generated by the PTE effect when thermalization is complete; (3) photocurrent immediately flows between the source-drain (S-D) contact without carrier transit time[8,12,25,36] through the Shockley-Ramo-type response[37,38]; (4) photocurrent decays via the carrier cooling, and a sample with lower mobility shows faster decay due to supercollision (SC) cooling[14,39]. Owing to the negligible RC time constant in our devices and the Shockley-Ramo-type response, the measured photocurrent directly tracks the development and evolution of the carrier temperature, allowing us to resolve multiple thermalization and cooling pathways that coexist in graphene[9,20,21,23,24]. The quantitative understanding of the above O-E conversion processes renews the basis for engineering ultrafast graphene optoelectronic devices.

We first present our experimental setup for ultrafast electrical readout. Then, we discuss the time for photovoltage generation and carrier transport in graphene PDs by measuring nonlocal photocurrent dynamics. Next, we describe the mechanism of photocurrent decay. Finally, we discuss the difference between our electrical readout and optoelectrical readout[5,12–15,17,40–42], which is commonly used to deduce the time scale of graphene PDs.

## Experimental setup

The basic concept of our experimental setup is illustrated in Fig. 1a. We made a high-quality single-layer-graphene transistor by hBN encapsulation with a ZnO top gate and Ti/Au edge contacts[43]. The ZnO gate is designed to be transparent in the gigahertz and terahertz range[34,35] so that the RC time constant due to the gate capacitance is negligible for ultrafast electrical readout (see Supplementary Section I-b for details). We fabricated four samples with different channel lengths ($L$ = 5 (two samples), 10, and 15 μm) and carrier mobilities in graphene ($\mu$ = 11,000 – 140,000 cm$^2$/Vs) (see Supplementary Section II-a for a detailed characterization). We refer to the four samples by their channel length: samples 5 μm[#1], 5 μm[#2], 10 μm, and 15 μm. As shown in the inset, one of the Ti/Au electrodes forming a Goubau-line waveguide is connected to a low-temperature-grown gallium arsenide (LT-GaAs) PC switch. Note that the Goubau line is suitable for investigating O-E conversion processes without the complexity of mode interference and mode coupling between graphene and the waveguide[30–32]. We carried out pump-probe experiments using a pulsed femtosecond laser with a 280-fs pulse duration. This beam was divided into pump and probe beams to excite the graphene and PC switch, respectively, with a controlled time delay to measure the photocurrent in the time domain. The pump beam was tightly focused on the graphene using an objective lens to perform scanning photocurrent microscopy[13,15,16,38]. We selected the wavelengths of 517 nm for better focus and a higher signal-to-noise ratio and 1035 nm for accurate Fermi-level tuning to avoid unwanted photo-induced doping[44] (see Supplementary Section II-b for details).

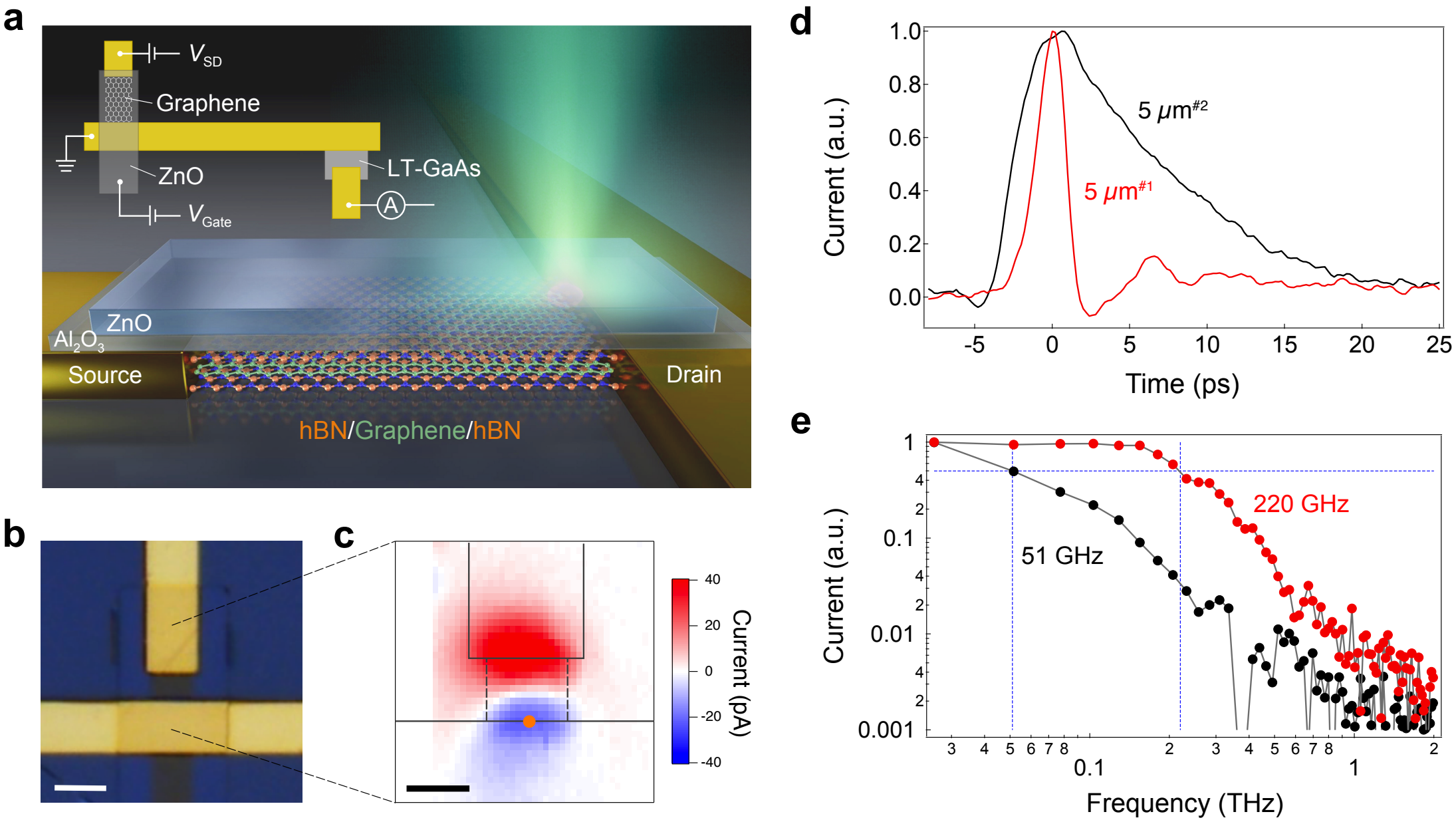

**Figure 1 | Setup for on-chip ultrafast electrical readout. a,** Schematic of the device structure. A ZnO top gate, which is designed to minimize the RC time constant, controls the Fermi level of hBN-encapsulated graphene. The ultrafast photocurrent is extracted from the drain contact with a Goubau line to measure its waveform at the LT-GaAs photoconductive switch using the pump-probe method. **b**, Optical image of the graphene device with 5-μm channel length. Scale bar: 10 μm. **c**, Scanning photocurrent image of **b**. Current is obtained at the peak of transient photocurrent. Scale bar: 5 μm. **d**, Temporal profiles of the photocurrent in samples 5 μm$^{\#1}$ and 5 μm$^{\#2}$ at the pump position indicated in **c** by the orange circle. For clarity, each waveform is normalized by the maximum peak current. Time origin is set at the current peak. **e**, Normalized Fourier-transformed spectra obtained from the waveforms in **d**. The 3-dB bandwidth is determined to be 220 and 51 GHz for samples 5 μm$^{\#1}$ and 5 μm$^{\#2}$, respectively. Experimental parameters: SD bias voltage $V_{SD}$ = 0 V; gate bias voltage $V_{Gate}$ = 0 V; laser wavelength $\lambda_{Laser}$ = 517 nm; pump power $P_{Laser}$ = 0.1 mW.

Figures 1b and c show an optical image of sample 5 μm[#2] and the corresponding scanning photocurrent image at zero SD bias voltage ($V_{SD}$), respectively. The photocurrent becomes maxima with opposite signs at the two graphene-metal interfaces, which is the signature of PTE current[16,38] (see Methods for details). Throughout the experiments, we kept $V_{SD}$ = 0 V to focus on O-E conversion by the PTE effect. Figure 1d shows time-resolved photocurrent for sample 5 μm[#1] ($\mu$ =11,000 cm$^2$/Vs) and sample 5 μm[#2] ($\mu$ =140,000 cm$^2$/Vs). The Fourier transform shows that the 3-dB bandwidth of sample 5 μm[#1] reaches 220 GHz (Fig. 1e), demonstrating that the cut-off frequency by the RC time constant is higher than 220 GHz as well as our system's capability to investigate ultrafast dynamics of O-E conversion processes. On the other hand, the bandwidth of sample 5 μm[#2], whose mobility is an order of magnitude higher than that of sample 5[#1], is as narrow as 51 GHz. This suggests that the carrier mobility has a strong impact on the bandwidth. In the following, we investigate the intrinsic mechanism behind the ultrafast O-E conversion by a comprehensive study, including its dependence on the Fermi level, mobility, S-D channel length, and pump-spot position.

## Experimental results

### Tunable thermalization time and instantaneous photocurrent response

First, we focus on the early stages of the O-E conversion, namely the generation of the photocurrent in the sample with $\mu$ = 140,000 cm$^2$/Vs ($L$ = 5 μm[#2]), the one with the highest mobility. Figure 2a shows the measured waveform for different gate voltages ($V_{Gate}$). The peak current varies with $V_{Gate}$ (inset of Fig. 2a) and is maximized around $E_F$ = -0.07 eV, which is consistent with the $V_{Gate}$ dependence of the Seebeck coefficient $S \propto d\sigma/dV_{Gate}$, where $\sigma$ is the graphene conductivity[3,45]. Most strikingly, as highlighted in Fig. 2b, the peak position of the photocurrent is shifted to a larger value of ~4 ps as the Fermi level is tuned towards the charge neutrality point (CNP). Note that negative photocurrent for $E_F$ ~0 is likely to be recently observed CNP photocurrent (see Supplementary Section III for details). It is tempting to explain the significant delay of the photocurrent generation by the carrier transit time in graphene[8,12,25,36], which is assumed to limit the bandwidth of graphene PDs. However, we verify below that the delay of the photocurrent generation is due to slow carrier thermalization time near the CNP, not to the carrier transit time in graphene.

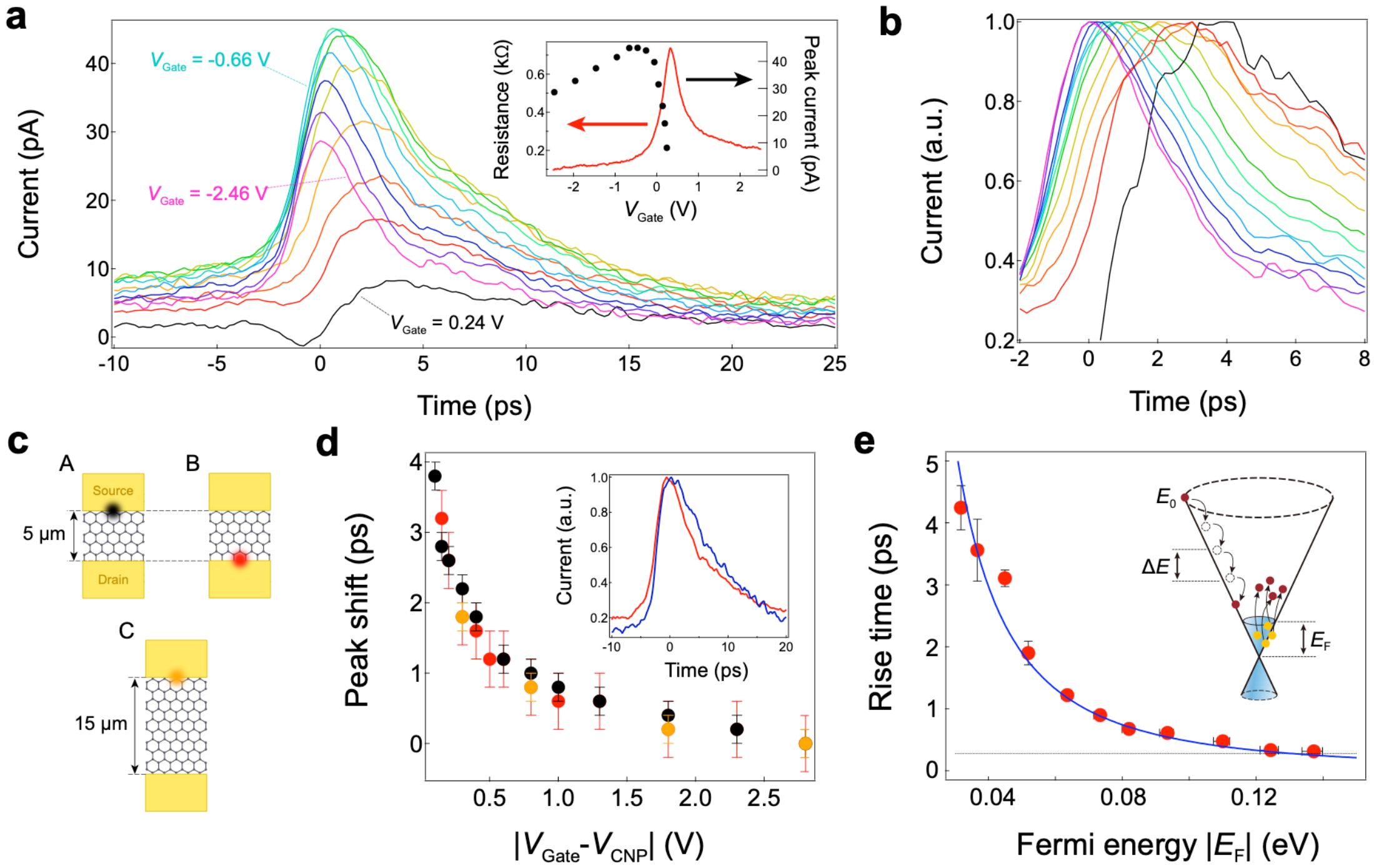

**Figure 2 | Instantaneous Shockley-Ramo response and tunable hot carrier multiplication.** **a**, Temporal profiles of the photocurrent in sample 5 μm[#2] with $\mu$ = 140,000 cm$^2$/Vs for several values of $V_{\mathrm{Gate}}$. Time origin is set at the current peak of $V_{\mathrm{Gate}}$ = -2.46 V. Inset shows corresponding peak current amplitude (black circles) and two-terminal resistance (red curve). **b**, Normalized waveform of **a** around the peak position, highlighting the gate-induced peak shift. **c**, Schematic of experimental configurations (A−C) for measuring the nonlocal photocurrent extraction dynamics. **a** is taken with configuration A. **d**, Peak shift extracted from each configuration (**c**), shown by corresponding color plots. These values show relative delay from $|V_{\mathrm{Gate}}\text{-}V_{\mathrm{CNP}}|$ = 2.8 V. Inset shows waveforms obtained with sample 10 μm by pumping upper (red) and lower (blue) graphene-metal interfaces ($V_{\mathrm{Gate}}$ = 0 V, $\lambda_{\mathrm{Laser}}$ = 517 nm). **e**, Rise time of the photocurrent as a function of the Fermi level. The blue curve shows the best fit obtained by the HCM model[11] with the adjustable scaling constant $\alpha$ = 7.9 ± 0.3 eV ps. The horizontal dotted line shows pulse duration of the pump pulse. Inset is a schematic of the HCM process. $\Delta E$ is the energy loss per step with $\Delta E \sim |E_{\mathrm{F}}|$. Error bars correspond to the estimated standard deviation of the fit coefficient. Experimental parameters are $V_{\mathrm{SD}}$ = 0 V, $\lambda_{\mathrm{Laser}}$ = 1035 nm, and $P_{\mathrm{Laser}}$ = 0.1 mW.

To uncover the cause of the delay, we evaluate the nonlocal photocurrent extraction dynamics by exciting different positions (interface to upper or lower contact) and the samples with different lengths and mobilities as schematically shown in Fig. 2c. The corresponding peak shift as a function of $V_{Gate}$ in the three configurations falls onto a single curve (Fig. 2d). This overlap indicates that the carrier transport in the graphene PD is not the cause of the delay, but instead is instantaneous (less than or comparable to the pump pulse duration of 280 fs) as showcased in the inset of Fig. 2d, in contrast to the naive expectation for the carrier transit time of ~18 ps for $L$ = 10 μm with the saturation velocity of $5.5 \times 10^5$ m/s[46]. This instantaneous response can be explained by the Shockley-Ramo (SR) theorem[37]. In conductive material like graphene, photocurrent is immediately generated when ambient carriers enter the contact, whereas photo-excited carriers themselves travel to the contact in a semiconductor. In the former case, the time scale of the photocurrent response is determined by the time for the screening of the long-range electric field. This SR response time is much shorter than the carrier transit time and is predicted to be around 300 fs for $L$ = 10 μm[37]. This indicates that the carrier transit time is independent of the bandwidth of a graphene PD, which is highly beneficial for developing ultrafast PDs.

Alternative possible cause for the delay is the time for the development of photovoltage induced by the thermalization into a Fermi-Dirac distribution with elevated carrier temperature. In graphene, energy dissipated to the phonon system does not contribute to the PTE current, since electronic heat capacity is orders of magnitude smaller than the lattice[3], and thus the rise time of photocurrent is determined by carrier-carrier scattering[9–11,15,21,22,38]. Since interband carrier scattering is minor due to the limited phase space[11,22,38], intraband carrier scattering dominates the thermalization dynamics except very close to the CNP[38] ($|E_F| <$ ~0.03 eV) (see Supplementary Section III for details). To quantitatively discuss the effects of intraband scattering on the thermalization dynamics, we determined the rise time of the photocurrent by an exponential rise-decay fitting as shown in Fig. 2e (see Supplementary Section IV for the detailed procedure). Above $|E_F| > 0.12$ eV, the rise time is determined by the duration of the pump pulse, while it increases with decreasing $|E_F|$ and exceeds 4 ps near the CNP. This slower dynamics can be explained by the smaller energy loss ($\Delta E$) per intraband scattering event for smaller $|E_F|$ during the thermalization. Namely, more scattering events are required to thermalize for smaller $|E_F|$ (inset of Fig. 2e) because the amount of energy exchanged between photo-excited carriers and carriers in the Fermi sea is $\Delta E \sim |E_F|$[11]. This cascade step leads to HCM with the number of secondary electron-hole pairs scaled by $E_0/|E_F|$, where $E_0$ is the excitation energy ($E_0/|E_F|$

>> 1). The HCM model[11], which suggests that the thermalization time is proportional to inverse square of the Fermi energy, $\Delta t = \alpha(E_0/E_\mathrm{F})/E_\mathrm{F}$ (with $\alpha = 7.9 \pm 0.3$ eV ps), well agrees with the experimental results. The tunable range of the rise time is an order of magnitude larger than that observed using intraband absorption of THz pulses[47] (from immediate to ~0.4 ps). This colossal tunability of the O-E conversion will be useful for ultrafast signal processing. We note that observing tunable photocurrent generation and identifying its origin become possible due to the capability of measuring nonlocal ultrafast photocurrent extraction with the negligible RC time constant, which is unique to our method.

**Mobility dependence of cooling**

Next, we discuss the photocurrent decay induced by the lowering of the carrier temperature. We note that, because of the PTE effect, SR response, and negligible RC time constant, the photocurrent directly tracks the time evolution of the carrier temperature. The responsivity of graphene PDs is known to be enlarged in a high-mobility device[5] due to the large Seebeck coefficient[48]. On the other hand, the mobility effect on the cooling time has not been quantitatively investigated. We show that the mobility is a crucial parameter for the cooling time adjustment by a systematic study using four samples with different mobilities at a common substrate and Fermi energy ($E_\mathrm{F} = -0.05$ eV) (Fig. 3a). Faster decay is observed for the samples with lower mobility. This behavior can be explained by the disorder-assisted SC cooling[14,39], where three-body collisions between carriers and both phonons and impurities take place. As shown in the bottom of Fig. 3b, the decay time in samples with mobility $\mu < 51{,}000$ cm$^2$/Vs (sample 15 μm) are reasonably explained by this SC prediction[14,23] (see Supplementary Section V for details). The decay time in sample 5 μm$^{\#2}$ with $\mu = 140{,}000$ cm$^2$/Vs is far below the SC prediction, suggesting that another pathway dominates the carrier cooling in this mobility regime. The $E_\mathrm{F}$ dependence of the decay time in this sample (see Fig. S7 in the Supplementary Information) suggests that the coupling to hyperbolic phonons in the hBN substrate[17] dominates at the high doping regime. At the low doping regime, where the decay is slow, the effects of the hyperbolic phonons are weak, and the optical phonons may be another major cooling pathway. The Seebeck coefficient calculated by the Mott formula[3,45] is also shown in the top of Fig. 3b to estimate the responsivity. We note that the obtained Seebeck coefficient up to 350 ± 40 μV/K is significantly high among various graphene PDs[2–9] owing to the high carrier mobility of our device by the hBN encapsulation. While there is a trade-off between bandwidth and responsivity, the large Seebeck coefficient with the first picosecond decay indicates the superior performance of hBN-encapsulated graphene. These results provide crucial information to tailor the performance of graphene PDs depending on their intended application.

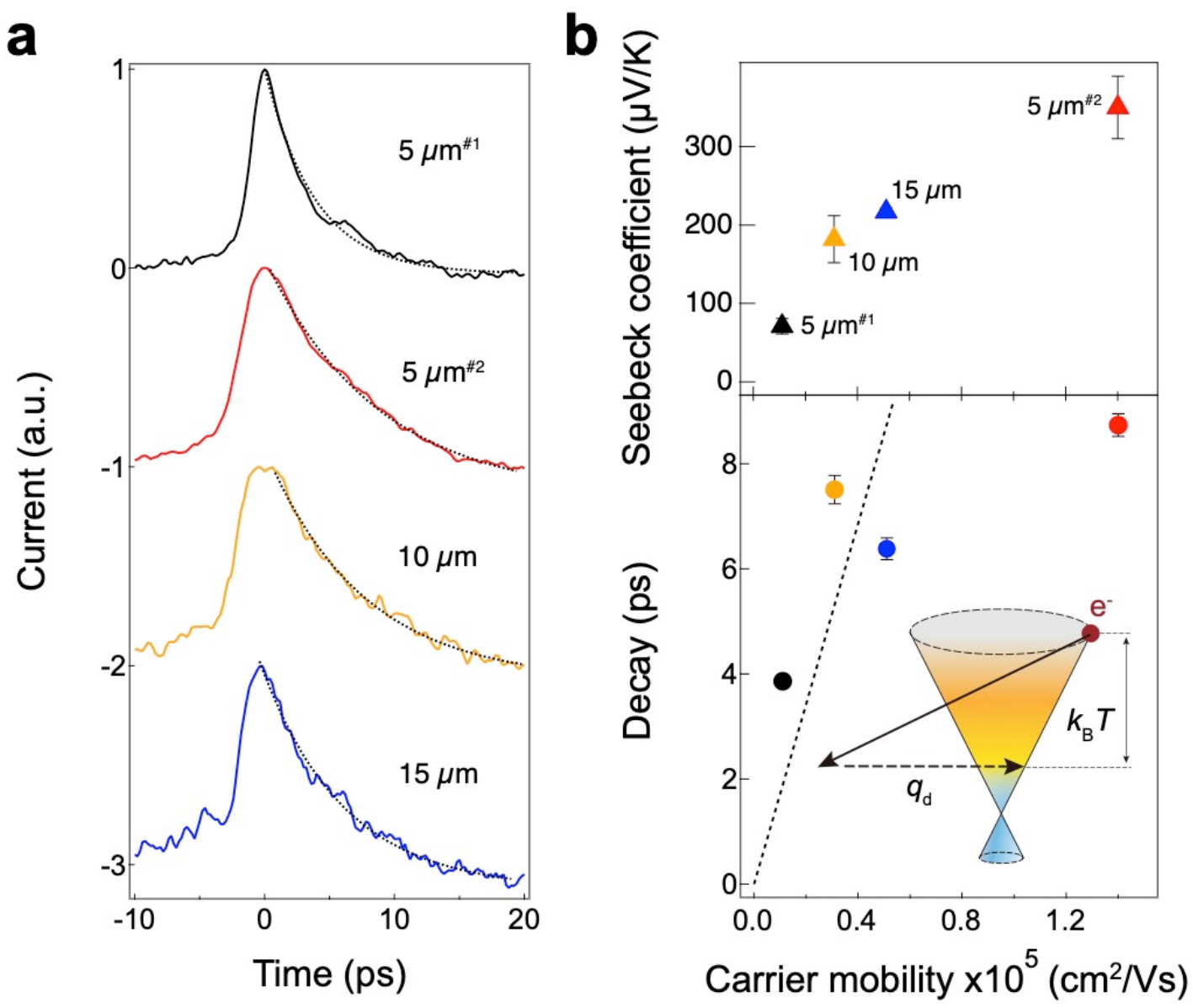

**Figure 3 | Cooling time in samples with different carrier mobilities. a**, Temporal profiles of the photocurrent obtained with different samples. For all the samples, the Fermi level is set to $E_F$ = -0.05 eV. Black dashed curves show the best fit by the exponential function. Traces are offset for clarity. **b**, Top figure shows the calculated Seebeck coefficient at $E_F$ = -0.05 eV (see Supplementary Section II-a for details). Bottom figure shows the decay constant obtained from **a** for the different samples as a function of the carrier mobility. Error bars correspond to the estimated standard deviation of the fit coefficient. The dashed line shows calculated SC cooling time with an adjustable parameter of the deformation potential $D$ = 27 eV (see Supplementary Section V for details). Inset shows the schematic of the SC cooling. Faster cooling by emission of high-energy ($k_BT$) acoustic phonons becomes possible due to relaxation of the momentum restriction by the disorder ($q_d$). Experimental parameters are $V_{SD}$ = 0 V, $\lambda_{Laser}$ = 1035 nm, and $P_{Laser}$ = 0.1 mW.

## Electrical readout vs optoelectrical readout of photocurrent decay

Finally, we further demonstrate the advantage of the electrical readout by comparing it with commonly used optoelectrical readout[5,12–15,17,40–42] (Fig. 4a), which is another method to measure the response of ultrafast photocurrent among various femtosecond pump-probe techniques[5,10,12–24]. The two different methods are schematically compared in Fig. 4a. In optoelectrical readout, graphene is excited by two laser pulses, and the photocurrent decay is measured as a function of the delay between them. Because photocurrent scales nonlinearly with the pump intensity, the photocurrent decreases relatively when two pump pulses overlap. Then, the decay constant is monitored as the recovery time of the photocurrent. While this local autocorrelation technique enables ultrafast time resolution up to sub 50 fs[15], it cannot measure nonlocal photocurrent dynamics such as the photocurrent extraction from the electrode. Moreover, evaluating the physical quantity deduced from the measured decay constant is challenging because it requires the assumption of the nonlinear response of the system, such as Pauli blocking[12], temperature-dependent electronic heat capacity[15,17], and Auger recombination[40], depending on experimental conditions.

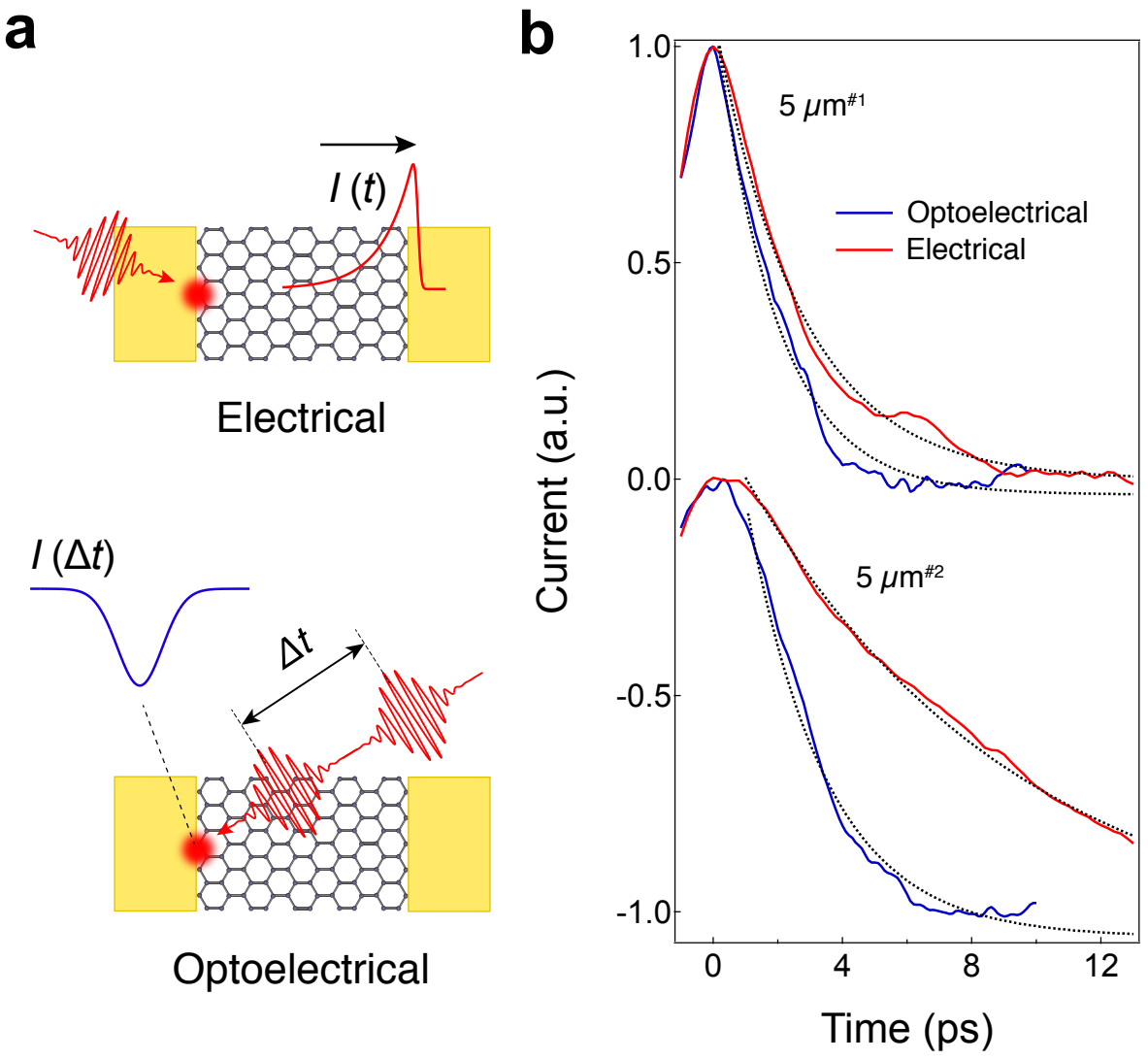

**Figure 4 | Electrical readout vs. optoelectrical readout. a**, Schematic of the two different methods. Electrical readout directly measures the ultrafast photocurrent extracted from the drain contact. Optoelectrical readout measures the local nonlinear response of the photocurrent at the focal spot as a function of the delay between two pump pulses. **b**, Temporal profiles of the photocurrent measured by the electrical and optoelectrical methods in samples 5 μm$^{\#1}$ with $\mu$ = 11,000 cm$^2$/Vs (upper traces) and 5 μm$^{\#2}$ with $\mu$ = 140,000 cm$^2$/Vs (lower traces). Black dashed curves show the best fits by the exponential function as guides for the eye. For clarity, each waveform is normalized by the maximum peak current and offset by 1. Experimental parameters are $V_{SD}$ = 0 V, $V_{Gate}$ = 0 V, $\lambda_{Laser}$ = 517 nm, and $P_{Laser}$ = 0.1 mW.

Compared to the time evolution of the carrier temperature tracked by electrical readout, optoelectrical readout on sample 5 μm[#2] gives a much faster decay (lower traces in Fig. 4b), demonstrating that the decay time measured by optoelectrical readout should be distinguished from the carrier cooling time in this experimental condition. We suggest that, among several origins of nonlinear optical response, Pauli blocking reflects decay of carrier occupation at the optical transition energy (1.2 eV). In defect-free graphene, microscopic first-principles calculation shows that, consistent with our observation, the decay of carrier occupation at visible wavelengths is much faster than that of carrier cooling time[20]. In graphene with lower mobility (sample 5 μm[#1] with $\mu$ =11,000 cm$^2$/Vs), on the other hand, the two readout methods give similar decay time (upper traces in Fig. 4b). We suggest that the SC cooling compensates for the difference between the decay of carrier occupation and carrier temperature because it can effectively relax electrons with energy lower than optical and hyperbolic phonons (~0.2 eV)[9]. These results demonstrate that direct readout of photocurrent is essential to understanding the O-E conversion processes and hence evaluating the performance of graphene PDs.

## Discussion

We demonstrated ultrafast nonlocal electrical readout of photocurrent in PTE-based graphene PDs. By combining on-chip THz spectroscopy and gate-tunable devices with a suppressed RC time constant, we succeeded in resolving O-E conversion processes in graphene with their intrinsic time scale. Contrary to common expectations, we found that the time for photocurrent generation is tunable from immediate at large Fermi level to > 4 ps at close to the CNP, determined by the thermalization of photoexcited carriers through tunable intraband HCM. We also found that the photocurrent response across the PDs is instantaneous without carrier transit time, following the SR theorem. Once the thermalization is attained, the photocurrent decays following the carrier temperature lowered through phonon interactions. Thanks to the immediate photocurrent flow and the negligible RC constant, our method works as ultrafast thermometry of graphene by directly tracking the time evolution of the carrier temperature. This enables us to obtain deep insights into the O-E conversion mechanism in graphene. Our prescriptions for overcoming the bandwidth limitations and gaining a quantitative understanding of the O-E conversion processes can be applied to any PTE-based graphene PD[2,4–9] and will set new standards for designing graphene optoelectronic devices. Furthermore, our platform developed here can be easily expanded for two-dimensional van der Waals materials and their heterostructures to investigate key ultrafast processes for O-E conversion enabled by the peculiar interlayer coupling, such as the interlayer hot carrier dynamics[49], interlayer excitons[50], and shift current[51]. Our on-chip ultrafast electrical readout removes the barrier between ultrafast optical science and device engineering by showing how to bring out the functionality (ultrafast O-E conversion) based on the fundamental understanding of ultrafast carrier dynamics, beneficial for the development of ultrafast optoelectronic applications.

## Methods

### Device fabrication

Photoconductive switches were prepared from an LT-GaAs wafer supplied form Batop GmbH. The wafer consists of a 2.6-μm-thick LT-GaAs surface layer (grown at 300°C) and a 500-nm-thick $Al_{0.9}Ga_{0.1}As$ sacrificial layer on a semi-insulating GaAs substrate[33]. After the LT-GaAs layer had been etched into 100 μm x 100 μm squares in a citric acid solution, the sacrificial layer was dissolved in hydrochloric acid. Etched LT-GaAs chips on the GaAs substrate were transferred to a sapphire substrate using a thermoplastic methacrylate copolymer (Elvacite 2552C, Lucite International) as an adhesive[52]. Remaining Elvacite on the sapphire substrate was removed in citric acid.

Graphene was prepared by mechanical exfoliation of natural graphite on $SiO_2$ (285 nm)/doped-Si substrates. Monolayer graphene was identified by optical contrast under a microscope. Using a separately exfoliated hexagonal boron-nitride (hBN) flake, graphene was picked up and transferred onto the sapphire substrate via the typical dry transfer technique with dimethylpolysiloxane (PDMS) and poly-carbonate (PC)[43].

Graphene was then patterned by reactive ion etching, and the Ti/Au waveguide structure with side contacts to graphene was sputtered. Then the whole surface was covered with the 30-nm-thick $Al_2O_3$ insulating layer grown by atomic layer deposition. The 20-nm-thick ZnO top gate grown by atomic layer deposition at 140°C was patterned on the $Al_2O_3$ layer by photolithography and liftoff processes. The ZnO top gate was protected by another $Al_2O_3$ layer on top. Finally, for electrical contact to the waveguide, $Al_2O_3$ on bonding pads was removed by Miroposit 351 developer.

### On-chip THz spectroscopy (electrical readout)

We employed a Coherent Monaco femtosecond laser as a light source (1035 nm, 280 fs, 16.8 MHz), and the second-harmonic wavelength of 517 nm was generated by Beta Barium Borate (BBO) crystal. We basically selected the 517-nm light for better focusing with the spot size of 1.8 μm (FWHM), but this wavelength inevitably fixes the Fermi level of the graphene near the CNP by photo-induced doping[44] (see Supplementary Section II-b for details); hence, we used 1035-nm light (spot size; 3.6 μm) when accurate Fermi level tuning was necessary. The two orthogonally polarized pump and probe beams were combined by using a polarization beam splitter and aligned with a slight displacement to focus them onto the graphene and LT-GaAs PC switch with an objective lens. The position of the pump beam was controlled by a motorized mirror, while the probe beam's position was kept constant throughout the experiments. An optical chopper modulated the pump beam at a few hundred hertz for lock-in detection of the THz current. All measurements were performed under ambient laboratory conditions.

The PTE photovoltage $V_{PTE}$ generated by pumping graphene-metal interface is given by

$$V_{PTE} = \Delta S(T_{el} - T_0), \quad (1)$$

where $\Delta S$ is the effective Seebeck coefficient, $T_{el}$ is the elevated carrier temperature, and $T_0$ is the initial temperature[16]. $\Delta S$ is maximized when pumping the graphene-metal interface because of discontinuity of the Fermi level due to metal-induced doping. The dynamics of $V_{PTE}$ and corresponding photocurrent that we measured in our devices are equal to the time evolution of $T_{el}$ thanks to the negligible RC time constant and carrier transit time.

### Autocorrelation of photocurrent (optoelectrical readout)

The identical measurement system to electrical readout (on-chip THz spectroscopy) is used for the optoelectrical readout. In the case of electrical readout, pump and probe beams excite the graphene and the PC switch, respectively, and the photocurrent is readout form the electrode connected to the PC switch. On the other hand, in the case of optoelectrical readout, pump and probe beams excite the same spot on the graphene (Fig.4a), and the photocurrent is readout from the electrode connected to the graphene. The polarization between two beams is set to perpendicular to remove unwanted interference when the two pulses overlap.

## Data availability

The datasets generated during and/or analysed during the current study are available from the corresponding author on reasonable request.

## Acknowledgements

The authors thank K. Sasaki and K. Nozaki for fruitful discussions, and H. Murofushi for technical support. K.W. and T.T. acknowledge support from the Elemental Strategy Initiative conducted by the MEXT, Japan (Grant Number JPMXP0112101001) and JSPS KAKENHI (Grant Numbers 19H05790, 20H00354 and 21H05233).

## Author contributions

K.Y. and N.K. conceived the experiment. K.Y. designed and built the optical setup, performed the measurement, and analysed the data. K.Y. and N.K. designed the THz circuits with support from T.W. and M.H. T.W. and N.K. fabricated devices. K.W. and T.T. contributed hBN material. K.Y. and N.K. wrote the paper, with input from all authors.

## Competing financial interests

The authors declare no competing financial interests.

Supplementary Information for

# Ultrafast intrinsic optical-to-electrical conversion dynamics in graphene photodetector

Katsumasa Yoshioka[1,*], Taro Wakamura[1], Masayuki Hashisaka[1], Kenji Watanabe[2], Takashi Taniguchi[3], and Norio Kumada[1]

[1]NTT Basic Research Laboratories, NTT Corporation, Atsugi, Japan
[2]Research Center for Functional Materials, National Institute for Materials Science, Tsukuba, Japan
[3]International Center for Materials Nanoarchitectonics, National Institute for Materials Science, Tsukuba, Japan
*e-mail: katsumasa.yoshioka.ch@hco.ntt.co.jp

## I-a. Sample design

Figure S1 shows the diagram of a THz circuit. The length of the 10-µm-wide and 200-nm-thick Ti/Au Goubau-line waveguide from the left bottom to right top bonding pads is about 24 mm, which is long enough so that the reflected photocurrent at bonding pads does not appear in the time window of interest. Graphene is electrically connected to the Goubau line and the top-left electrode. The distance between graphene and the LT-GaAs PC switch is about 550 µm, slightly sample dependent due to the difference in the graphene transfer position. Connection to the ZnO gate was made by silver past at the end of the electrode.

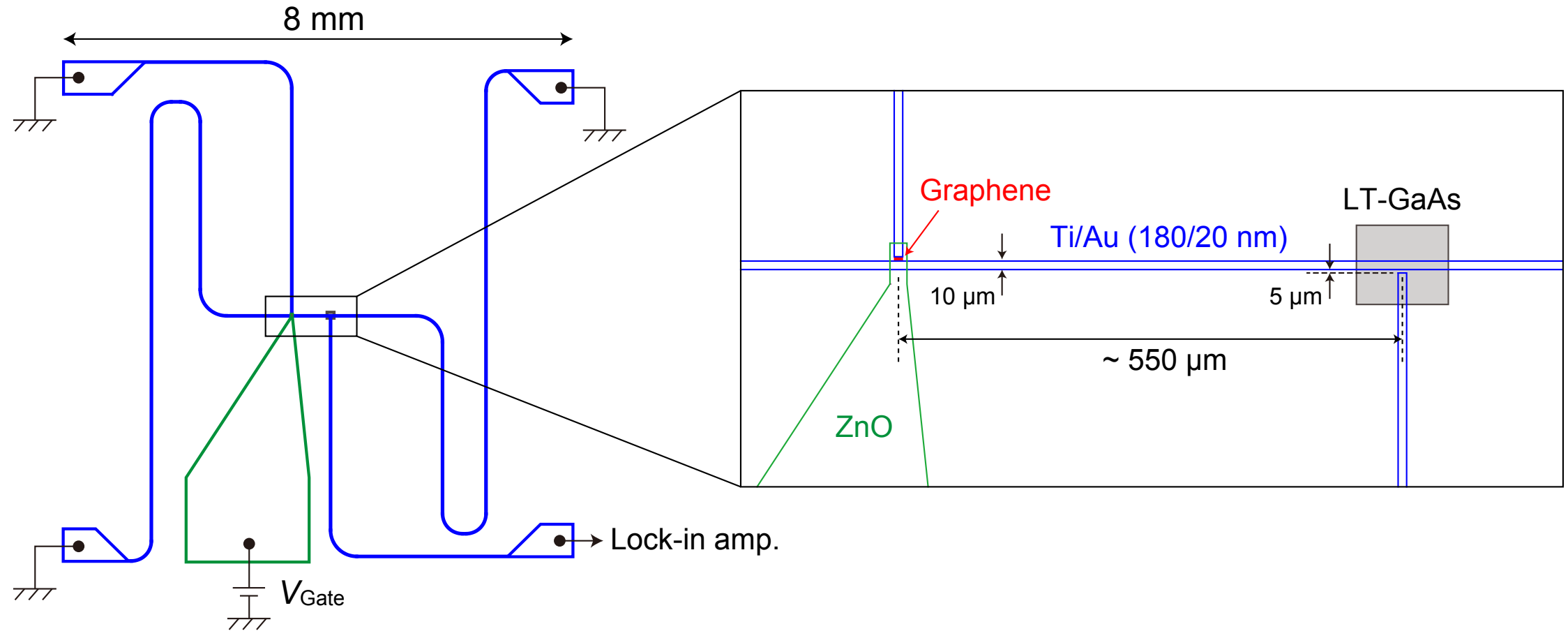

**Figure S1 | Diagram of the THz circuit and the electrical connection for the current measurement.**

## I-b. Properties of the ZnO film

The RC time constant of our device is dominated by $R_{con}C_{gate}$[1], where $R_{con}$ is the contact resistance, and $C_{gate}$ is the gate capacitance. However, effective $C_{gate}$ can be suppressed by decreasing the conductivity $G$ of the gate material. When $G$ is low enough, charge carriers in the gate electrode cannot follow a high-frequency electric field and thus the material behaves like an insulator at the high frequency.

To find an appropriate condition, we tested several ZnO films with different $G$ values, which can be controlled by the growth temperature. Figure S2 shows the transmittance of 20-nm-thick ZnO films measured by Fourier transform infrared spectroscopy (FTIR) at 5 THz as a function of the ZnO conductivity; higher transmittance indicates less screening of the THz electric field. We found that the transmittance becomes larger than 90% when $G \lesssim 5000$ $Sm^{-1}$. We also found that carrier density in graphene can be controlled properly by DC bias using these ZnO films, except for the one with the lowest conductivity $G$ = 100 $Sm^{-1}$, as a gate. As a result, we used ZnO films with $1000 < G < 5000$ $Sm^{-1}$ in this work. We have confirmed transparency of the ZnO film in our two previous experiments: FTIR for graphene plasmons in the 1.5 – 7 THz range[2] and high-frequency transport for plasmons in InAs two-dimensional electron systems in the 1 – 10 GHz range[3]. Both experiments have demonstrated that, in samples with a ZnO gate, high-frequency response is the same as ungated samples while the carrier density is still tunable by a DC bias. These results ensure that the RC time constant originating from the top gate is negligible in the present frequency range. Remaining $C$ is the one between the two electrodes connected to graphene. We confirmed that this capacitance is negligibly small (<1 fF) by a finite element method (COMSOL Multiphysics).

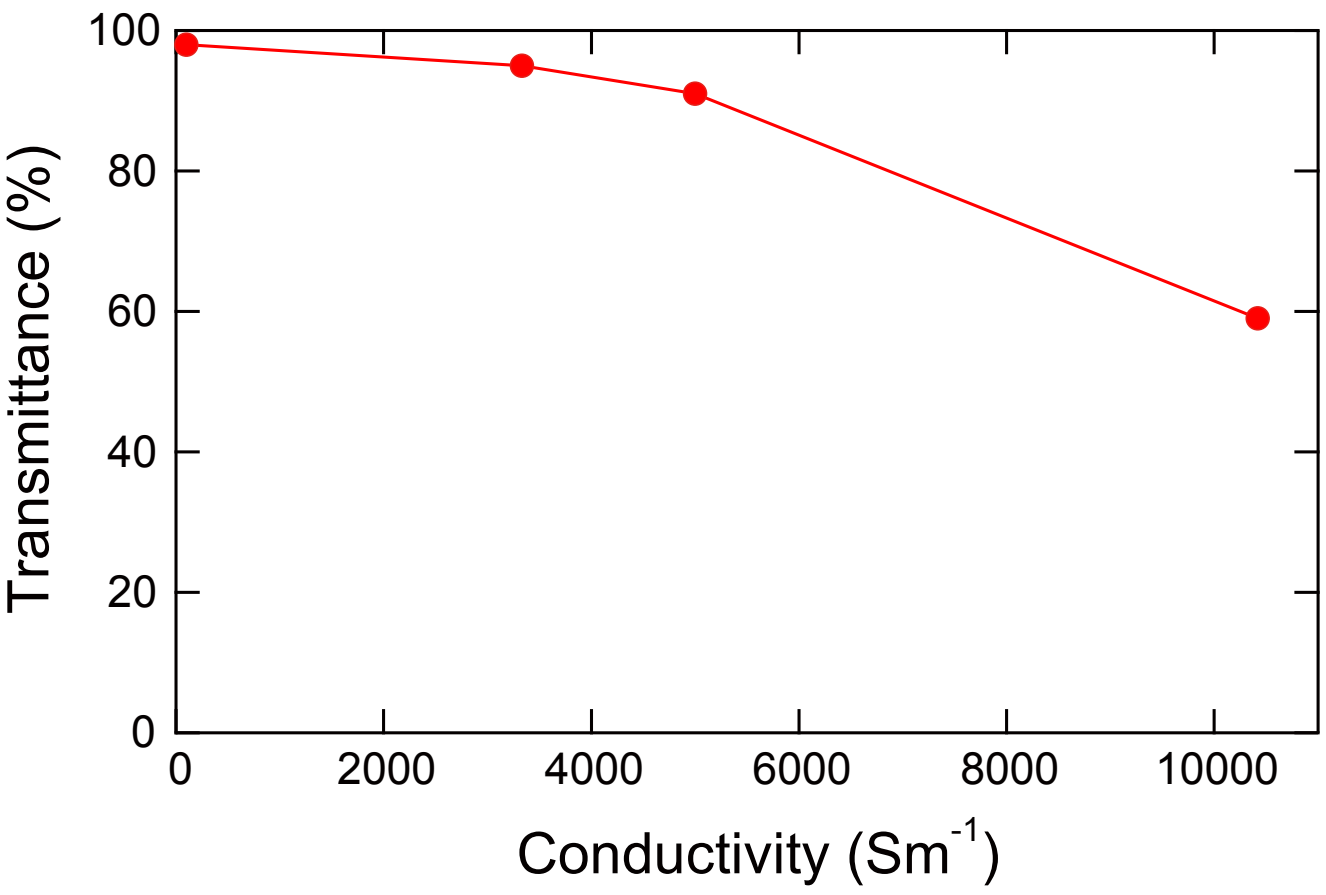

**Figure S2 | Transmittance of 20-nm-thick ZnO films at 5 THz as a function of the ZnO conductivity.**

### II-a. Sample characterization

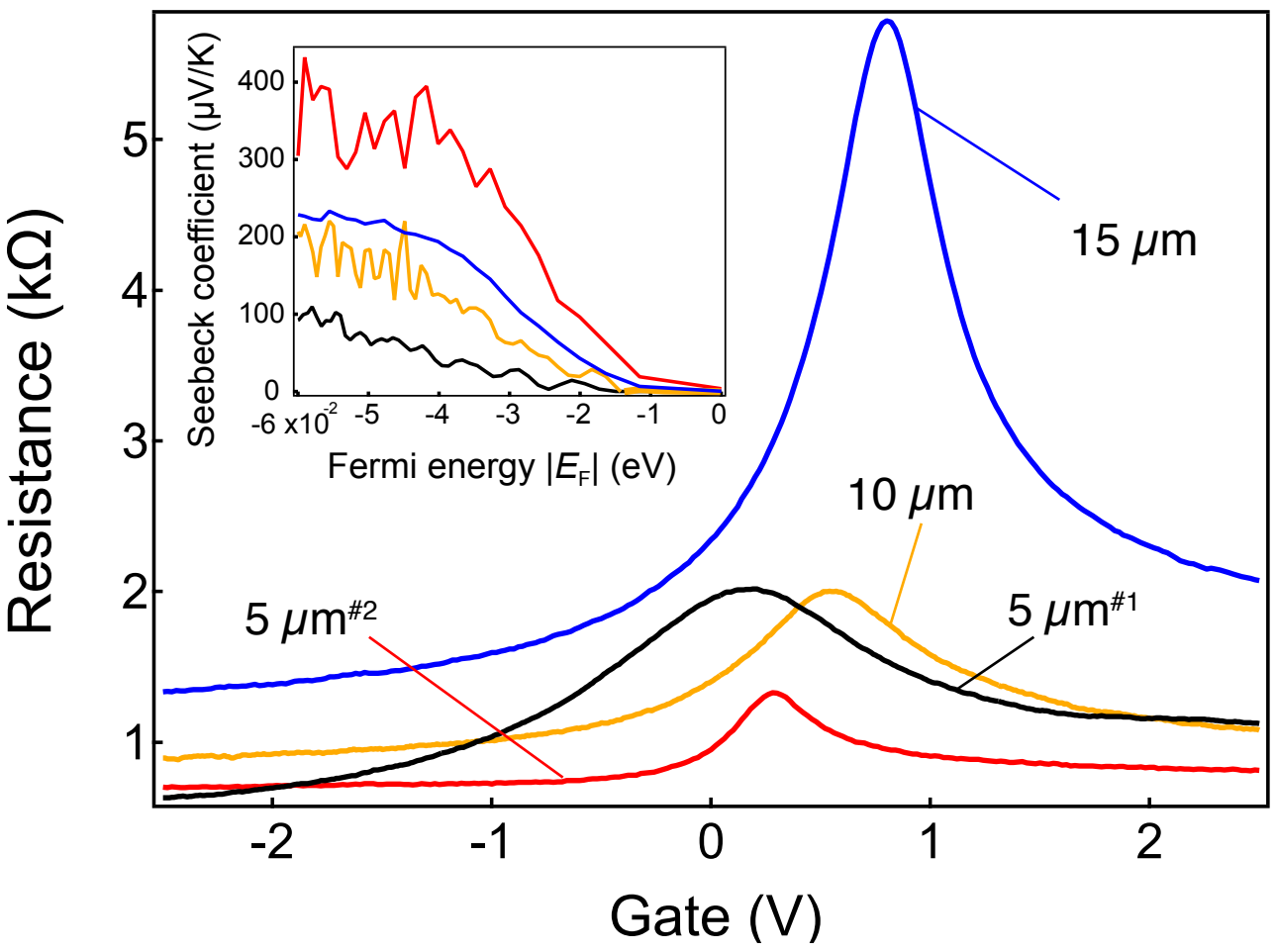

**Figure S3 | Two-terminal resistance of the graphene devices as a function of $V_{\mathrm{Gate}}$.**

| Sample | 5 μm#1 | 5 μm#2 | 10 μm | 15 μm |
| --- | --- | --- | --- | --- |
| Mobility ($cm^2$/Vs) | 1,1000 | 140,000 | 31,000 | 51,000 |
| Contact resistance (Ω) | 180 | 80 | 210 | 530 |
| Capacitance (fF) | 35 | 32 | 63 | 83 |
| Seebeck coefficient ($E_F$ = -0.05 eV) (μV/K) | 70 ± 10 | 180 ± 30 | 220 ± 5 | 350 ± 40 |

**Table S1 | Key parameters for each device.**

Figure S3 shows the two-terminal resistance of fabricated devices with different channel lengths. The carrier mobility $\mu$ is estimated from the gradient of sheet conductance in the *p*-doped regime in accordance with photocurrent measurements, as $\mu = \Delta\sigma_s/(\Delta V_{\mathrm{gate}} C_g)$, where $C_g$ is the gate capacitance. The key parameters for each device are summarized in Table S1. Note that $C_g$ becomes zero in a high-frequency regime (> GHz) because of the highly resistive zinc oxide top gate[2,3]. All data were taken in ambient laboratory conditions. We used a sourcemeter (2400, KEITHLEY) for the DC characterization, while a lock-in amplifier (LI 5640, NF Corporation) was used for the ultrafast readout of photocurrent based on the pump-probe scheme. Inset shows the calculated Seebeck coefficient using the Mott formula[4,5] from the two-terminal resistance. The Fermi energy is estimated by $E_F = \pm\hbar v_F \sqrt{\pi n}$ ,where $v_F$ is the Fermi velocity and $n$ is the carrier density given by $n = C_{Gate}\left(V_{Gate} - V_{CNP}\right)/e$, where $e$ is the elementary charge.

### II-b. Photo-induced doping in hBN encapsulated graphene

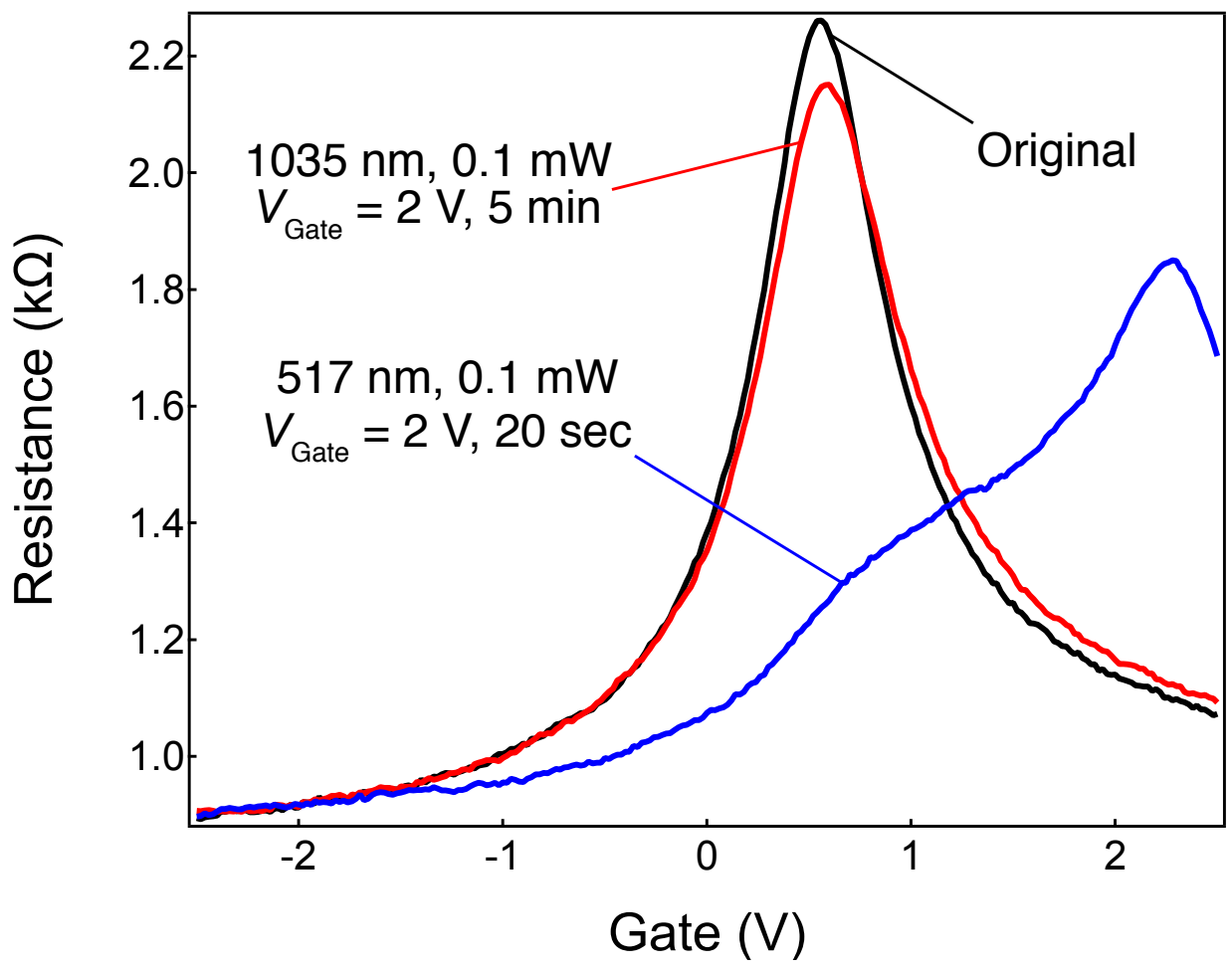

**Figure S4 | Effect of photo-induced doping in graphene.**

We investigate the photo-induced modulation doping effect in graphene, which is particularly strong in graphene/hBN heterostructures[6]. Photo-induced doping inevitably shifts the Fermi energy to around the charge neutrality point (CNP) regardless of where $V_{Gate}$ is set because ionized defects in hBN created by photoexcitation effectively screen the electric field induced by the top gate. In other words, the doping level under illumination is determined by the $V_{Gate}$. Figure S4 shows the two-terminal resistance of the 10-μm sample upon light illumination. The original state (black curve) is obtained after homogeneous light irradiation with $V_{Gate} = 0$ V, $\lambda_{Laser} = 517$ nm, and $P_{Laser} = 0.1$ mW for over ten minutes until the graphene is stabilized. This procedure enables us to initialize graphene with homogeneous doping all over the sample. The blue curve is obtained after 20-s light irradiation with $V_{Gate} = 2$ V. As expected, the CNP drastically shifts to around 2 V with broadened shape because of inhomogeneous doping coming from the Gaussian distribution of light intensity. This situation is unacceptable for a quantitative understanding of O-E conversion processes. In contrast, irradiation by 1035-nm light shows a negligible CNP shift (red curve) even with a long irradiation time of 5 min. This insensitivity suggests that defect levels of hBN are deeper than 1.2 eV (1035 nm) but accessible with 2.4 eV (517 nm).

Therefore, to maintain the homogeneous doping in graphene, we kept $V_{Gate}$ = 0 V for 517-nm light excitation (Fig. 1c-e; Fig. 2d inset; Fig 4b in the main text), while we used 1035-nm light when Fermi level control was required (Fig. 2 a-e; Fig. 3 in the main text). Note that we checked the two-terminal resistance before and after experiments and confirmed that the inhomogeneous doping was absent during the measurements throughout the paper.

### III. Interband electron-electron scattering

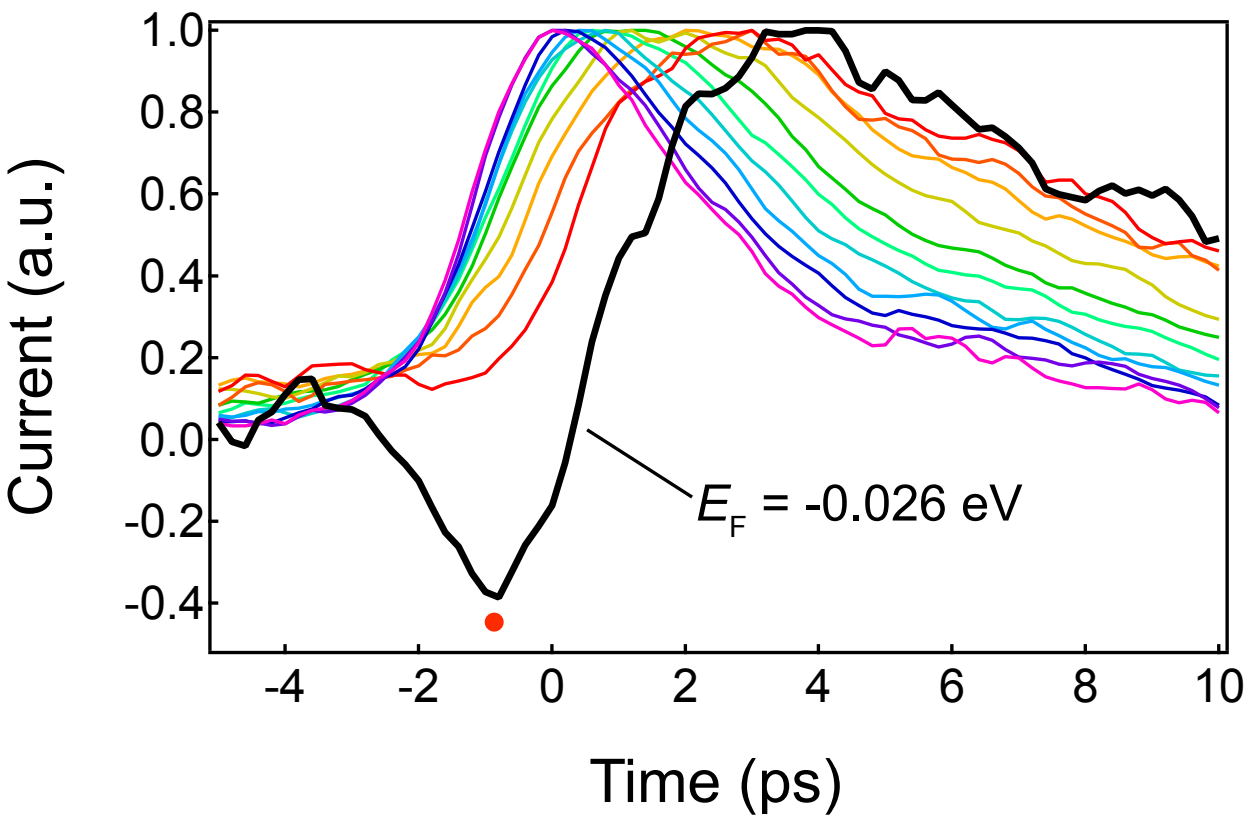

**Figure S5 | Instantaneous photocurrent generation near charge neutrality point.**

As described in the main text, photocurrent generation time is determined by Fermi-level-dependent intraband scattering in graphene. However, as highlighted by the red circle in Fig. S5, anomalous negative photocurrent kicks in when the Fermi level is tuned very close to the CNP. In contrast to the delayed positive PTE current, negative current shows an instantaneous response. We believe that this immediate photocurrent is coming from the recently discovered CNP photocurrent[7]. At the CNP, interband scattering in a colinear direction is dominant because of the limited phase space due to the energy and momentum conservation. The colinear scattering preserves the initial photocurrent immediately created by photoexcitation due to different group velocities of electrons and holes at elevated energies. This CNP photocurrent is predicted to generate without waiting for thermalization, which agrees well with our observation. Note that alternative explanation might be possible by discontinuity of the Fermi level near the contact due to metal-induced doping.

## IV. Estimation of rise time

For a quantitative understanding of the thermalization process, we determined the rise time of the photocurrent (Fig. 2e in the main text) using the exponential rise and decay function given by $A\cdot\left(e^{-t/\tau_{\mathrm{decay}}} - e^{-t/\tau_{\mathrm{rise}}}\right)$ convoluted with a Gaussian-shaped system response function with the FWHM of 1.8 ps determined by the optoelectrical readout of photocurrent at GaAs PC switch. Figure S6 shows the best fit to the experimental data (Fig. 2b in the main text) with different $V_{\mathrm{Gate}}$, and extracted $\tau_{\mathrm{rise}}$ values are shown in Fig. 2e. As expected, the obtained rise time shows a similar value to the peak shift, shown in Fig. 2d in the main text.

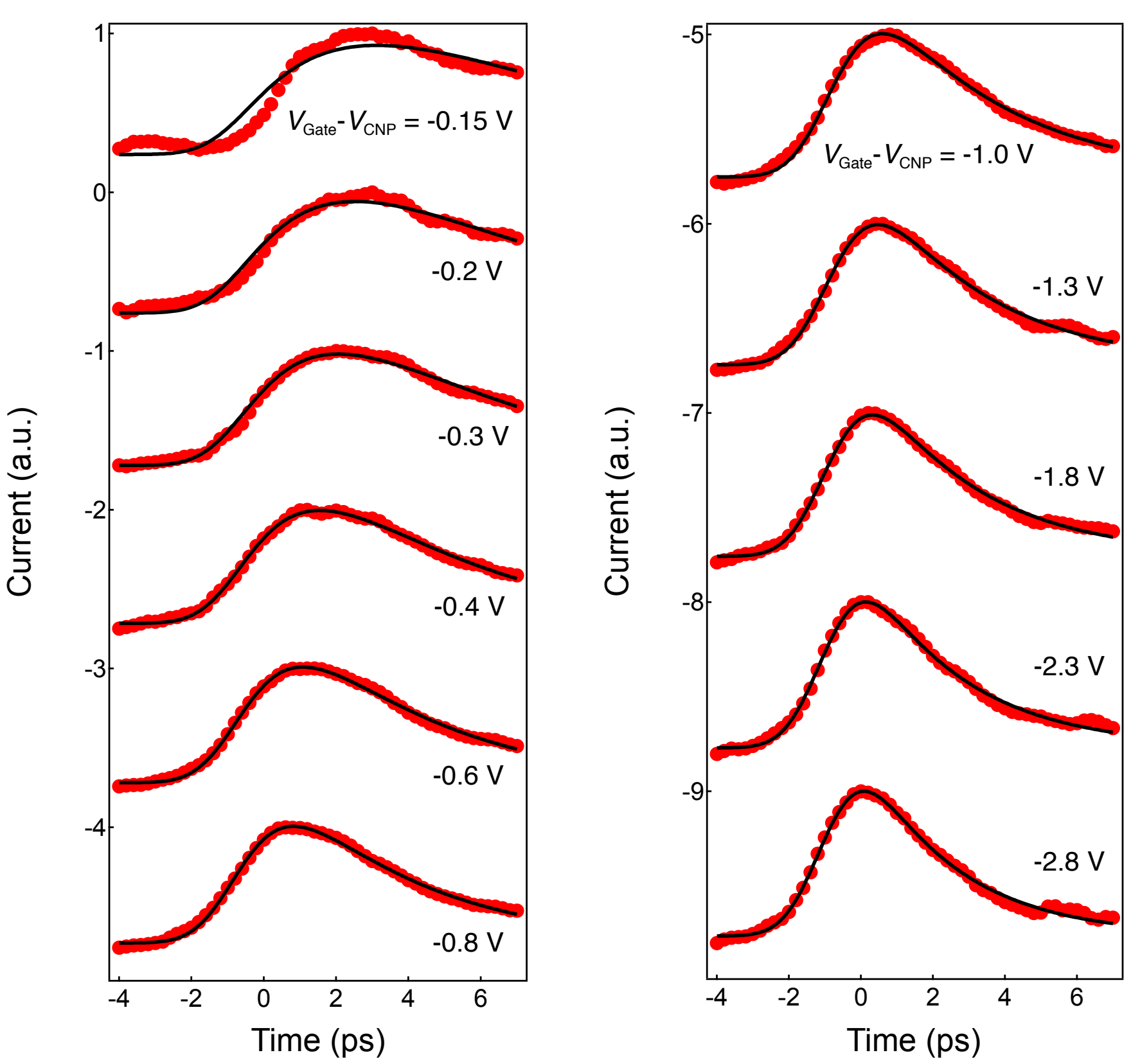

**Figure S6 | Estimation of rise time using exponential rise and decay function.**

## V. Supercollision cooling in graphene

To explain the mobility dependence of the photocurrent decay time (Fig. 3b in the main text), we calculated cooling time through disorder-assisted scattering with acoustic phonons[8,9] (or supercollision (SC) cooling) as $\tau_{\mathrm{sc}} = \left(3\frac{A}{\alpha}\cdot T_l\right)^{-1}$, where $T_l$ is the lattice temperature (300 K), and $\frac{A}{\alpha} = \frac{2\lambda k_{\mathrm{B}}}{3k_{\mathrm{F}} l \hbar}$. Here, $k_{\mathrm{B}}$ is Boltzmann's constant, $k_{\mathrm{F}} = \frac{E_{\mathrm{F}}}{\hbar \upsilon_{\mathrm{F}}}$ is the Fermi momentum with $\upsilon_{\mathrm{F}}$ the Fermi velocity, and $l = \upsilon_{\mathrm{F}}\tau_{\mathrm{ms}}$ is the mean free path with $\tau_{\mathrm{ms}} = \frac{\mu E_{\mathrm{F}}}{e\upsilon_{\mathrm{F}}^2}$ the momentum scattering time. Finally, $\lambda = \frac{2D^2 E_{\mathrm{F}}}{\rho \upsilon_{\mathrm{s}}^2 \pi \hbar^2 \upsilon_{\mathrm{F}}^2}$ is the electron-phonon coupling strength, where $D$ is the deformation potential, $\rho$ is the mass density, and $\upsilon_{\mathrm{s}}$ is the sound velocity. We used $E_{\mathrm{F}} = 0.05$ eV, $\upsilon_{\mathrm{F}} = 10^6$ m/s, $\rho = 7.6 \times 10^{-7}$ kg/m$^2$, and $\upsilon_{\mathrm{s}} = 2.0 \times 10^4$ m/s[10] for the calculation. As shown in Fig. 3b, samples with mobility below $\mu < 51{,}000$ cm$^2$/Vs ($L$ = 15 μm) are reasonably explained by this SC cooling with $D = 27$ eV. This value agrees with a previous experimental study ($D < 35$ eV) conducted using hBN encapsulated graphene[11].

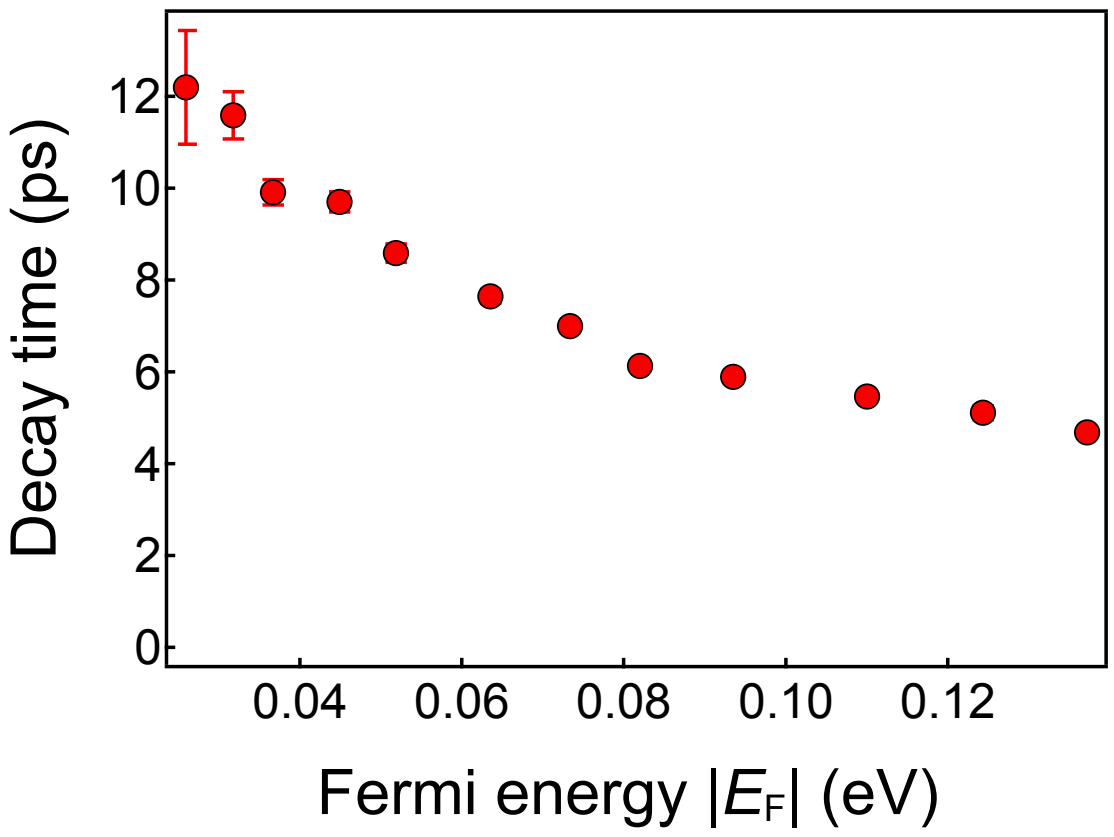

**Figure S7 | $E_{\mathrm{F}}$ dependence of the decay time in sample 5 μm[#2] with $\mu$ =140,000.** Decay constant obtained from Fig. 2a in the main text. Error bars correspond to the estimated standard deviation of the fit coefficient. Faster cooling for the larger $E_{\mathrm{F}}$ is the signature of the coupling to hyperbolic phonons in the hBN substrate[11].

## VI. Response time of LT-GaAs photoconductive switch

To determine the response time of the LT-GaAs PC switch, we fabricate the THz circuit with two PC switches connected with 550-μm-long Goubau line following Ref. 12. Figure S8 shows obtained time-domain waveform and corresponding fit with the symmetrical response function given by[12]

$$I_{\text{meas}}(t) \propto \frac{1}{2} e^{\frac{\sigma^2 - t\tau}{\tau^2}} \pi \sigma^2 \tau \left( \text{Erfc}\left[ \frac{-t + \frac{2\sigma^2}{\tau}}{2\sqrt{\sigma^2}} \right] + e^{\frac{2t}{\tau}} \text{Erfc}\left[ \frac{2\sigma^2 + t\tau}{2\sqrt{\sigma^2}\,\tau} \right] \right),$$

where $\sigma^2$ is the variance of a Gaussian function connected to the laser pulse FWHM by $\tau_{\text{Laser}} = 2\sqrt{2\ln(2)}\sigma$, and $\tau$ is the lifetime of GaAs photoexcited carriers. As a result, the response time $\tau$ is determined to be $1.80 \pm 0.03$ ps.

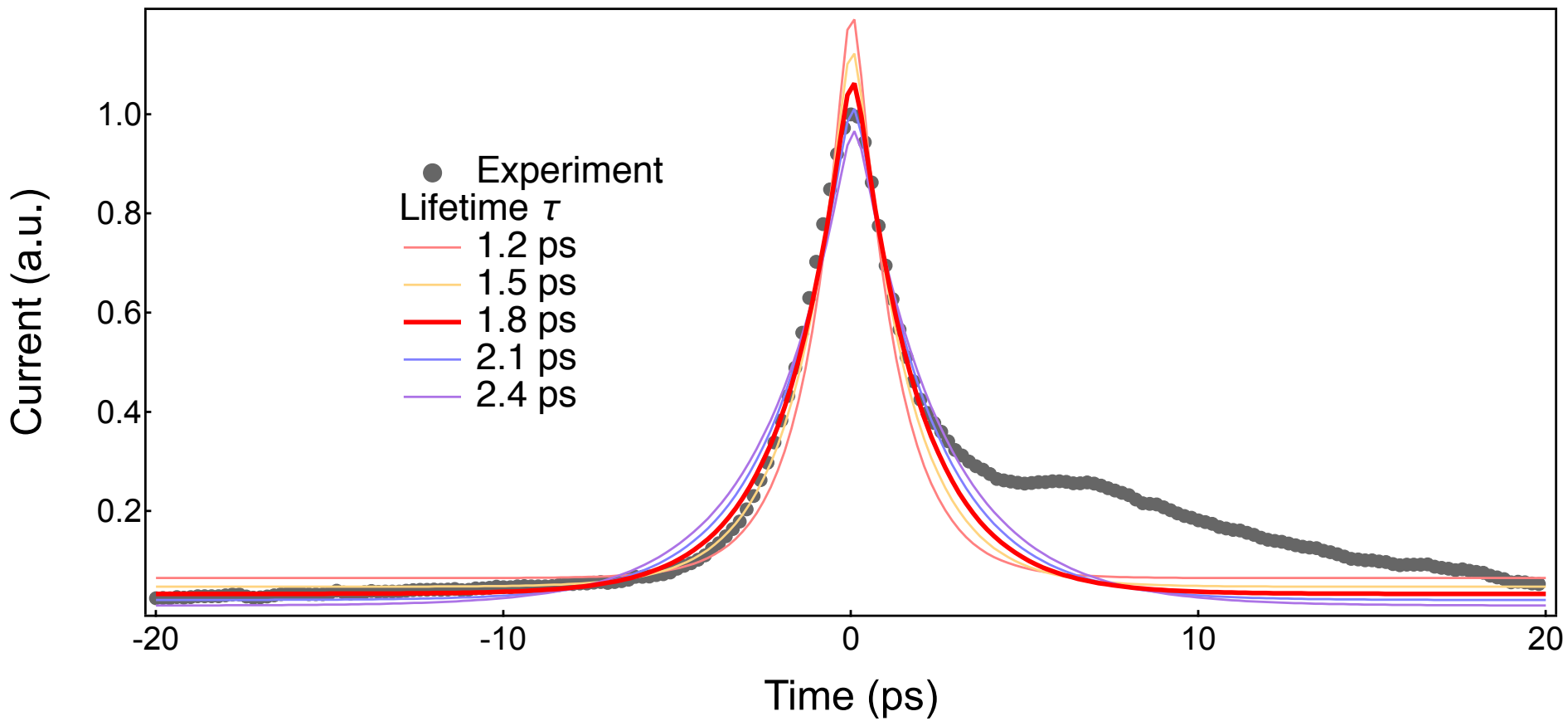

**Figure S8 | Estimation of the response time of the PC switch.** The predetermined pulse duration of $\tau_{\text{Laser}} = 280$ fs is used for the fitting. The best fit is obtained with the response time of $\tau = 1.80 \pm 0.03$ ps. Fitted results with other $\tau$ is also shown for the clarity. The right shoulder from ~4 ps stems from a reflection at the detection switch and a back reflection at the generation PC switch.